\documentclass[twocolumn,english,aps,prl,showpacs,superscriptaddress]{revtex4-1}
\usepackage[T1]{fontenc}
\usepackage[latin9]{inputenc}
\usepackage{amsmath}
\usepackage{amssymb}
\usepackage{graphicx}

\makeatletter
\@ifundefined{textcolor}{}
{%
 \definecolor{BLACK}{gray}{0}
 \definecolor{WHITE}{gray}{1}
 \definecolor{RED}{rgb}{1,0,0}
 \definecolor{GREEN}{rgb}{0,1,0}
 \definecolor{BLUE}{rgb}{0,0,1}
 \definecolor{CYAN}{cmyk}{1,0,0,0}
 \definecolor{MAGENTA}{cmyk}{0,1,0,0}
 \definecolor{YELLOW}{cmyk}{0,0,1,0}
}

\makeatother

\usepackage{babel}
\begin{document}

\title{Hole propagation in the orbital compass models}

\author{Wojciech Brzezicki }
\email{w.brzezicki@uj.edu.pl}
\affiliation{Marian Smoluchowski Institute of Physics, Jagellonian University,
Reymonta 4, PL-30059 Krak\'ow, Poland }
\affiliation{Dipartimento di Fisica 'E.R. Caianiello', Universita di Salerno,
I-84084 Fisciano (Salerno), Italy}

\author{Maria Daghofer}
\affiliation{Institut f\"ur Theoretische Festk\"orperphysik, IFW Dresden,
D-01171 Dresden, Germany}

\author{Andrzej M. Ole\'{s} }
\affiliation{Marian Smoluchowski Institute of Physics, Jagellonian University,
Reymonta 4, PL-30059 Krak\'ow, Poland }
\affiliation{ Max-Planck-Institut f\"ur Festk\"orperforschung, Heisenbergstrasse 1,
D-70569 Stuttgart, Germany }

\date{15 May, 2014}
\begin{abstract}
We explore the propagation of a single hole in the generalized
quantum compass model which interpolates between fully isotropic
antiferromagnetic (AF) phase in the Ising model and nematic order
of decoupled AF chains for frustrated compass interactions.
We observe coherent hole motion due to either interorbital hopping
or due to the three-site effective hopping, while quantum spin
fluctuations in the ordered background do not play any role.
\end{abstract}

\pacs{75.10.Jm, 03.65.Ud, 64.70.Tg, 75.25.Dk}

\maketitle

Properties of strongly correlated transition metal oxides are
determined by effective interactions in form of spin-orbital
superexchange, introduced first long ago by Kugel and Khomskii
\cite{Kug82}. The spin-orbital interactions have enhanced quantum
fluctuations \cite{Kha05} and are characterized by frustration and
entanglement \cite{Ole12}. It leads, for instance, to rather cute
topological order in an exactly solvable SU(2)$\otimes XY$ ring
\cite{Brz14}. To understand better the consequences of directional
orbital interactions, it is of interest to investigate doped orbital
systems \cite{Wro10}.

Probably the simplest model that describes orbital-like superexchange
is the two-dimensional (2D) orbital compass model (OCM) \cite{vdB13}.
The so-called generalized compass model (GCM) introduced later
\cite{Cin10} provides a possibility to investigate a second order
quantum phase transition (QPT) between the Ising model and generic OCM
when frustration increases. The orbital anisotropies are captured in
the OCM with different spin components coupled along each bond,
$J_x\sigma_{i}^{x}\sigma_{j}^{x}$ and
$J_z\sigma_{i}^{z}\sigma_{j}^{z}$ along $a$ and $b$ axis of the
square lattice. Recent interest in this model is motivated by its
interdisciplinary character as it plays a role in the variety of
phenomena beyond the correlated oxides:
($i$) it is dual to recently studied models of $p+ip$ superconducting
arrays \cite{Nus05},
($ii$) it provides an effective description for Josephson arrays of
protected qubits \cite{Dou05} realized in recent experiments
\cite{Gla09}, and
($iii$) it could also describe polar molecules in optical lattices
\cite{Mil07}, as well as nitrogen-vacancy centers in a diamond
matrix \cite{Tro12}.

An exact solution of the one-dimensional (1D) generalized variant of
the compass model \cite{Brz07} gives a QPT at $J_{x}=J_{z}$.
A similar QCP occurs in the 2D OCM between types of 1D nematic orders:
for $J_{x}>J_{z}$ ($J_{x}<J_{z}$), antiferromagnetic (AF) chains form
along $a$ ($b$) that are --- in the thermodynamic limit --- decoupled
along $b$ ($a$). It has been shown that the symmetry allows one to 
reduce the original $L\times L$ compass cluster to a smaller 
$(L-1)\times(L-1)$ one with modified interactions which made it 
possible to obtain the full exact spectra and the specific heat for 
larger clusters \cite{Brz10}.
Electron itinerancy has been addressed in the weak-coupling limit
at temperatures above the ordering transition \cite{Nus12}.

In this paper we will discuss the motion of a single hole in the
ordered phases of the GCM, including the nematic phases of the simple
OCM. Following \cite{BrzDa}, we obtain the spectral functions of the 
itinerant models that reproduce GCM in the strong coupling regime. 
A great advantage of using the itinerant models is that a variational 
cluster approach (VCA) can be used to obtain unbiased results for both 
weak and strong coupling regime. The VCA was introduced to study 
strongly correlated electrons in models with local interactions 
\cite{Dah04,Pot03}. Since the interactions are here Ising-like, 
quantum fluctuations are suppressed and the paradigm for hole 
propagation known from the spin $t$-$J$ model does
no longer apply. This happens for the $t_{2g}$ electrons in $ab$
planes of Sr$_{2}$VO$_{4}$ where instead holes move mostly via
three-site terms \cite{Dag08,Woh08}. In case of $e_{g}$ electrons,
interorbital hopping delocalizes holes within ferromagnetic LaMnO$_3$
planes \cite{vdB00}.
In the present case, hole propagation occurs through quantum processes
involving the hole itself, rather than those of the ordered background.

The 2D GCM with AF interactions ($J>0$) is,
\begin{equation}
{\cal H}_{J}^{\theta}=J\sum_{i}\left\{
 \bar{\sigma}_{i}(\theta)\bar{\sigma}_{i+a}(\theta)
+\bar{\sigma}_{i}(-\theta)\bar{\sigma}_{i+b}(-\theta)\right\},
\label{eq:H_gcmp}
\end{equation}
with $\bar{\sigma}_{i}(\theta)$ being the composed pseudospins,
\begin{equation}
\bar{\sigma}_{i}(\theta)=
\cos(\theta/2)\sigma_{i}^{x}+\sin(\theta/2)\sigma_{i}^{z},
\label{eq:sigb}
\end{equation}
interpolating between $\sigma_{i}^{x}$ for $\theta=0$ and
$(\sigma_{i}^{x}\pm\sigma_{i}^{z})/\sqrt{2}$ for $\theta=\pi/2$ and
$\{\sigma_{i}^{x},\sigma_{i}^{z}\}$ are $S=1/2$ pseudospin operators.
$\{i+a(b)\}$ is a shorthand notation for the nearest neighbor of site
$i$ along the axis $a(b)$. For $\theta=0$ GCM corresponds to the
classical Ising model with $S^x_i$ components coupled on all the bonds.
In the opposite limit $\theta=\pi/2$ describes the OCM in a rotated
spin space: bonds along $a$ couple the spin component $(S^x_i+S^z_i)$
and bonds along $b$ the orthogonal one $(S^x_i-S^z_i)$. For
$0<\theta<\pi/2$, the GCM interpolates between Ising and compass
models \cite{Cin10}. The rotation of orbital operators (\ref{eq:sigb})
provides a convenient way to detect the phase transition between 2D
Ising and nematic compass order: In the former, moments lie along $x$
while they lie along either $x+z$ (in the following identified with
lattice axis $a$) or $x-z$ in the latter.

GCM follows from the two-orbital Hubbard model \cite{BrzDa},
\begin{eqnarray}
{\cal H}_{t-U} &=& t\sum_{i}\sum_{{\mu,\nu=\atop \alpha,\beta}}\left\{
      A_{\mu\nu}c_{i,\mu}^{\dagger}c_{i+a,\nu}^{}
\!+\! B_{\mu\nu}c_{i,\mu}^{\dagger}c_{i+b,\nu}^{}
\!+\!{\rm H.c.}\right\}\nonumber \\
&+& U\sum_{i}n_{i,\alpha}n_{i,\beta},
\label{eq:H_tU}
\end{eqnarray}
at large $U$ and half filling, where $A_{\mu,\nu}$ and $B_{\mu,\nu}$
are hopping matrices in $a$, $b$ directions between orbitals $\alpha$
and $\beta$. These can be obtained using standard perturbation theory
for two neighboring sites as,
\begin{eqnarray}
A_{\theta} &\!=\!&\frac{1}{\sqrt{2}}\left(\begin{array}{cc}
1+\sin\frac{\theta}{2} & \cos\frac{\theta}{2}\\
\cos\frac{\theta}{2} & 1-\sin\frac{\theta}{2}
\end{array}\right)\!=\!\frac{1}{\sqrt{2}}\bigl[1+\bar{\sigma}(\theta)\bigr],
\label{eq:Ahop_gcmp}\\
B_{\theta} &\!=\!&\frac{1}{\sqrt{2}}\left(\begin{array}{cc}
1+\sin\frac{\theta}{2} & -\cos\frac{\theta}{2}\\
-\cos\frac{\theta}{2} & 1-\sin\frac{\theta}{2}
\end{array}\right)\!=\!\frac{1}{\sqrt{2}}\bigl[1-\bar{\sigma}(-\theta)\bigr].
\label{eq:Bhop_gcmp}
\end{eqnarray}
The pseudospins $\{\sigma_{i}^{x},\sigma_{i}^{z}\}$ are the quadratic
forms of the fermions $c_{i}^{\dagger}$, i.e.,
$\sigma_{i}^{z}=n_{i,\alpha}- n_{i,\beta}$,
$\sigma_{i}^{x}=c_{i,\alpha}^{\dagger}c_{i,\beta}
+ c_{i,\beta}^{\dagger}c_{i,\alpha}$
and the superexchange is $J=t^{2}/U$.
In the small-$U$ regime the properties of the itinerant model of Eq.
(\ref{eq:H_tU}) can be well described by a mean-field (MF) approach
\cite{BrzDa}.

\begin{figure}[t!]
\begin{minipage}[t]{1\columnwidth}%
\begin{center}
\includegraphics[clip,width=7.8cm]{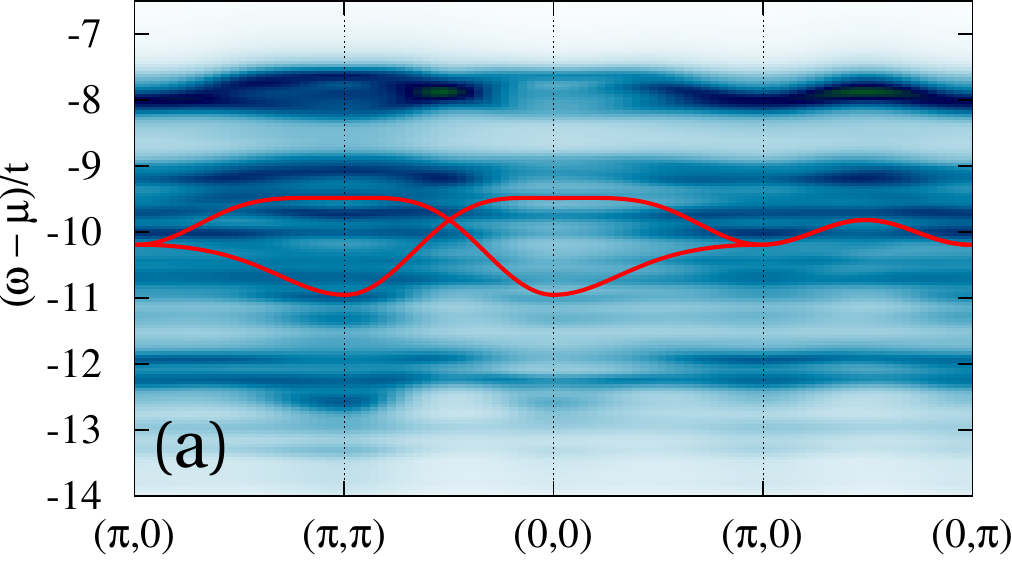}
\par\end{center}
\end{minipage}
\begin{minipage}[t]{1\columnwidth}%
\begin{center}
\includegraphics[clip,width=7.8cm]{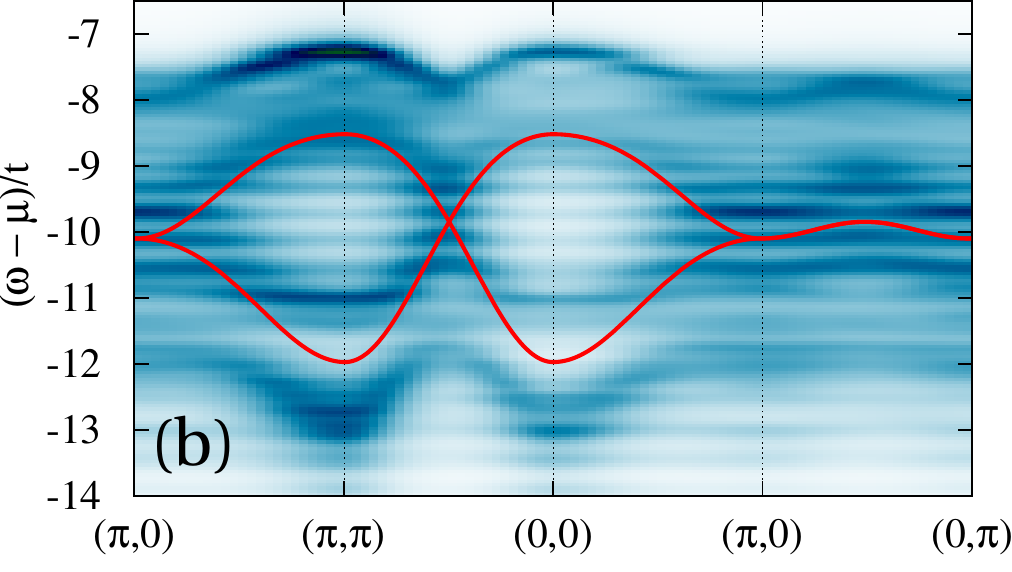}
\par\end{center}
\end{minipage}
\begin{minipage}[t]{1\columnwidth}%
\begin{center}
\includegraphics[clip,width=7.8cm]{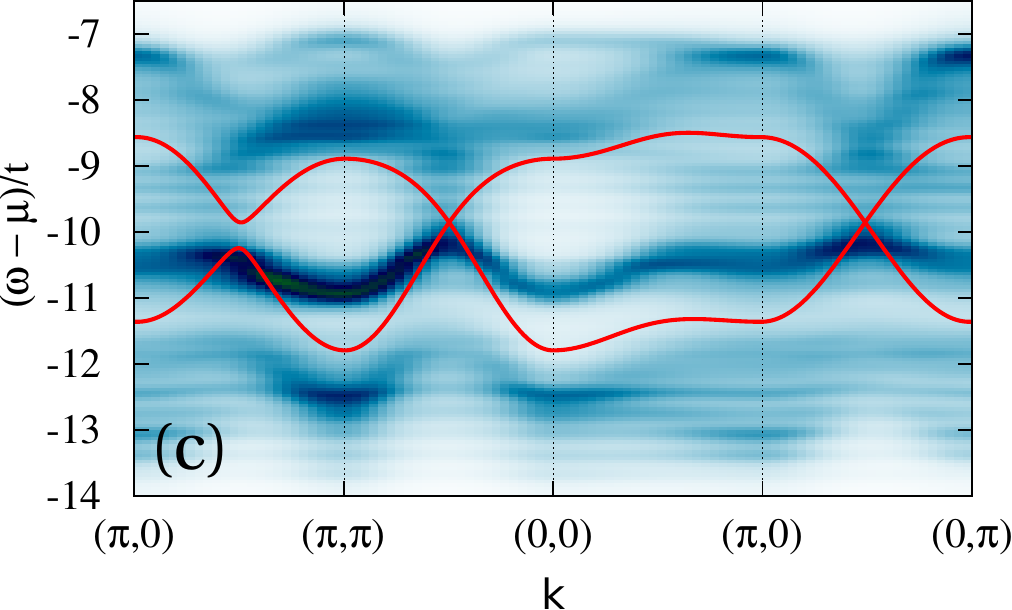}
\par\end{center}
\end{minipage}
\caption{Spectral functions obtained in the VCA at strong coupling 
($U=20t$) for the GCM with increasing frustration of interactions at:
(a) $\theta=\pi/6$,
(b) $\theta=5\pi/12$, and
(c) $\theta=89^{\circ}$.
The plots (a) and (b) refer to the AF$x$ phase of GCM and plot (c) to
the AF$a$ one ($\theta_c^{{\rm VCA}}\approx 88^{\circ}$). Solid lines
stand for the MF bands. }
\label{fig:spec_gcmp}
\end{figure}

Let us first discuss in more depth the somewhat surprising result that
a hole does not couple the AF chains of the OCM. The relevant are the 
row/column flips along the $x$ or $z$ axis. To see their impact on the 
itinerant model it is more convenient to look at the OCM in its 
original basis at site $i$, $\{\tau_{i}^{z},\tau_{i}^{x}\}$.
Setting $\tau_{i}^{z,x}=(\sigma_{i}^{x}\pm\sigma_{i}^{z})/\sqrt{2}$
we can easily transform GCM at $\theta=\pi/2$ into OCM with
$\tau_{i}^{z}\tau_{i+a}^{z}$ bonds along the $a$ axis and
$\tau_{i}^{x}\tau_{i+b}^{x}$ along the $b$ one.
Now we can see that OCM commutes with
$P_{i}=\prod_{n}\tau_{i+nb}^{z}$ and $Q_{i}=\prod_{n}\tau_{i+na}^{x}$
operators and the hopping matrices take form of
\begin{equation}
\label{hop}
A_{0}=\left(\begin{array}{cc}
1 & 0\\
0 & 0
\end{array}\right),\quad
B_{0}=\frac{1}{2}\left(\begin{array}{cc}
1 & 1\\
1 & 1
\end{array}\right).
\end{equation}
To see the action of the row flips in the fermion space, the operator
$Q_{i}$ should be first generalized to the case of double and zero
occupancy of site $i$. This is achieved by modifying $\tau_{i}^{x}$ as
follows,
$\tau_{i}^{x}\to\tilde{\tau}_{i}^{x}\!=\!(1-n_{i})^{2}+\tau_{i}^{x}$,
so that $(\tilde{\tau}_{i}^{x})^{2}=1$.
Now we can produce new $\tilde{Q}_{i}$ operator in the same way as
before and see its action on the fermion operators, which is
$\tilde{Q}_i(c_{j,\alpha(\beta)})\tilde{Q}_{i}=c_{j,\beta(\alpha)}$,
for all $c_{j,\mu}$ lying on the line of $\tilde{Q}_{i}$ and unity
for the others. Under this change the interaction part of the
${\cal H}_{t-U}$ remains unchanged, i.e.,
$\tilde{Q}_i{\cal H}_U\tilde{Q}_i\!=\!U\sum_{i}n_{i,\alpha}n_{i,\beta}$.
After a single row-flip $B_{0}$ remains invariant and $A_{0}$ changes as
\begin{equation}
A_{0}=\left(\begin{array}{cc}
1 & 0\\
0 & 0
\end{array}\right)\to\left(\begin{array}{cc}
0 & 0\\
0 & 1
\end{array}\right).
\end{equation}
This brings us to the conclusion that ${\cal H}_{t-U}^{0}$ is covariant
under the action of the $\tilde{Q}_i$; the exact form of the 
Hamiltonian changes, but the change is such that the properties of the
new Hamiltonian are the same as before --- only the orbitals along one
line are renamed which is irrelevant for the physics. This is important
for the VCA calculation as it allows us to calculate one-particle
spectra in one of the nematic ground states, e.g., the AF one, instead
of having to average over many of them~\cite{BrzDa}. For the OCM, 
results were tested for finite-size effects by changing cluster 
geometry and size; data presented here are for a $3\times4$ cluster.

In what follows we compare results obtained by the VCA and by MF. 
A first difference concerns the critical angle $\theta_{c}$ of the QPT 
from the AF$x$ phase to the AF$a$ one: Whereas the VCA value 
$\theta_c^{{\rm VCA}}\approx 88^{\circ}$ is close to the quasi-exact
$\theta_c^{\textrm{{\rm MERA}}}\simeq 84.8^{\circ}$ \cite{Cin10}, MF 
deviates more strongly with $\theta_{c}^{{\rm MF}}\approx 68^{\circ}$. 
While such a discrepancy might suggest the importance of quantum 
fluctuations within the AF background, we are going to see that 
processes related to the hole itself are more important.

Figure \ref{fig:spec_gcmp} illustrates how the spectral density changes 
across the QPT from the Ising to nematic order for increasing $\theta$.
For $\theta=\pi/6$, see Fig. \ref{fig:spec_gcmp}(a), which is very 
close to the classical AF Ising model, we see a ladder spectrum typical 
for the $\theta=0$ limit, because at weak quantum fluctuations the hole 
is confined in a string potential. The two MF bands can naturally not 
reflect such a ladder spectrum. Nevertheless, MF bands reflect the 
limited hole mobility and thereby qualitatively reproduce the shape
of the topmost VCA band.

For $\theta=5\pi/12$, see Fig. \ref{fig:spec_gcmp}(b), the bands become 
significantly more dispersive, especially the ones on the top. The shape 
of the topmost band  continues to be qualitatively well reproduced by 
the MF and this band is the sharpest feature seen in the spectral 
function at $\theta=5\pi/12$. We observe that the bands predicted by MF 
repel each other in the VCA and new features emerge at the intermediate
energies, with rather incoherent weight. Similarly to the generic OCM
case, see Fig. \ref{fig:spec_gcmp}(c) in \cite{BrzDa}, bands are most
dispersive along the direction $(0,0)\to(\pi,\pi)$. Since the ground 
state is still Ising ordered (AF$x$ phase), the increased dispersion, 
especially of the rather coherent topmost band, is here not primarily 
driven by quantum fluctuations. Instead, interorbital hopping is now 
significant, see Eqs. (\ref{eq:Ahop_gcmp}) and (\ref{eq:Bhop_gcmp}), 
which allows the hole to propagate,
similar to the case of a hole in $e_{g}$ orbital order \cite{vdB00}.

Finally, in Fig. \ref{fig:spec_gcmp}(c) we show the spectral function
at $\theta>\theta_c$ in the AF$a$ nematic order ($\theta=89^{\circ}$).
The bottom band is seen as a coherent feature which roughly agrees with
the MF prediction, but is much less dispersive. The upper band cannot
be identified so easily, even though the features around
$\vec{k}=\left(\pi/2,\pi/2\right)$ resemble the MF bands. Strong
coupling differences to the MF bands are on one hand the incoherent
weight and on the other the separation of bottom and top bands. One
finds that the MF bands do not really cross at 
$\vec{k}=\left(\pi/2,\pi/2\right)$, but they remain very close to each 
other. In the VCA, they are better separated, suggesting a strong 
effective interaction at this value of $\vec{k}$ that cannot be 
captured by a simple MF approach. A distinct feature observed in Fig. 
\ref{fig:spec_gcmp}(c) is a rather coherent band in the middle of the 
spectrum, absent in the MF approach. It seems to strongly repel the 
two bands at the top and bottom of the spectrum, thus making them 
flatter and widening the overall spectrum. We have shown that 
three-site hopping is the mechanism responsible for the observed 
dispersion of this additional band \cite{BrzDa}.

Summarizing, we have seen that the coherent motion of a single hole
(present for any $\theta>0$) is due to:
($i$) interorbital hopping in the AF phase, and 
($ii$) three-site hopping for the nematic order. 
MF cannot fully describe either case, it misses the ladder spectrum 
due to the string potential (AF order) and the three-site hopping 
(nematic order). In both cases, motion is thus due the quantum 
fluctuations caused by the hole itself rather than by the 
fluctuations of the ordered background.

\acknowledgments

W.B. acknowledges the kind hospitality of the Leibniz Institute for
Solid State and Materials Research in Dresden.
W.B. and A.M.O. acknowledge support by the Polish National
Science Center (NCN) under Project No. 2012/04/A/ST3/00331.
M.D. thanks Deutsche Forschungsgemeinschaft
(grant No. DA 1235/1-1 under Emmy-Noether Program) for support.

\end{document}